\documentclass[twocolumn,twoside,slac_two]{revtex4}
\usepackage{amsmath}
\usepackage{graphicx}
\usepackage{fancyhdr}
\usepackage{multirow}
\newcommand{\mev}{\ensuremath{\mathrm{MeV}}}
\newcommand{\gev}{\ensuremath{\mathrm{GeV}}}
\newcommand{\fb}{\ensuremath{\mathrm{fb}}}
\newcommand{\ifb}{\ensuremath{\mathrm{fb}^{-1}}}

\newcommand{\elec}{\ensuremath{\mathrm{e}^-}}
\newcommand{\pos}{\ensuremath{\mathrm{e}^+}}
\newcommand{\epem}{\pos\elec}
\newcommand{\dmes}{\ensuremath{\mathrm{D}}}
\newcommand{\dbar}{\ensuremath{\overline{\mathrm{D}}}}
\newcommand{\dstar}{\ensuremath{\mathrm{D}^\ast}}
\newcommand{\dstarp}{\ensuremath{\mathrm{D}^{\ast+}}}
\newcommand{\dstarb}{\ensuremath{\overline{\mathrm{D}}{}^\ast}}
\newcommand{\dplus}{\ensuremath{\mathrm{D}^+}}
\newcommand{\dz}{\ensuremath{\mathrm{D}^0}}
\newcommand{\BBbar}{\ensuremath{\mathrm{B}\overline{\mathrm{B}}}}

\newcommand{\ccbar}{\ensuremath{c\overline{c}}}
\newcommand{\ccres}{\ensuremath{(\ccbar)_{res}}}
\newcommand{\ufours}{\ensuremath{\Upsilon(4S)}}

\newcommand{\aprod}{\ensuremath{\alpha_{\text{prod}}}}
\newcommand{\ahel}{\ensuremath{\alpha_{\text{hel}}}}
\newcommand{\tprod}{\ensuremath{\theta_{\text{prod}}}}
\newcommand{\thel}{\ensuremath{\theta_{\text{hel}}}}
\newcommand{\br}{\ensuremath{\mathcal{B}}}

\newcommand{\mrecoil}{\ensuremath{M_{\text{recoil}}}}

\begin{document}

\title{Double \ccbar\ production in \epem\ annihilations at high energy}

\author{B.D.~Yabsley}
\affiliation{School of Physics, University of Sydney. NSW 2006, AUSTRALIA.}

\begin{abstract}
We review the current state of experimental knowledge on double \ccbar\
production in \epem\ annihilation.
The large cross-sections ($\mathcal{O}(20\,\fb)$)
for $\epem\to\gamma^*\to\psi^{(\prime)}\,\ccres$ processes
have been confirmed by detailed tests and reproduced by a second group:
they should now be considered well-established.
The latest experimental results concern the case
where the second \ccbar\ system is above open-charm threshold:
hidden-charm states continue to play a prominent role in the mass spectrum.
Some ``loose ends'' in the field are also briefly discussed. 
\end{abstract}

\maketitle

\thispagestyle{fancy}

\section{Introduction}

Electron-positron annihilation to charmonium and an additional hidden-charm state,
or a pair of open-charmed mesons --- double \ccbar\ production --- has been
known for only five years, but is now an established object of theoretical
and experimental study. The field is still being driven by data.
This presentation surveys current experimental knowledge:
after reviewing the history (Section~\ref{sec:history}) and 
the now-established results on double-charmonium production
$\epem\to\psi^{(\prime)}\,\ccres$ (Section~\ref{sec:baseline}),
we discuss the latest experimental work,
which focusses on states above open-charm threshold
(Section~\ref{sec:cutting}).
Some ``loose ends'', which might reward renewed attention,
are also noted (Section~\ref{sec:sidelines}).
In closing (Section~\ref{sec:summary})
we summarize both established and new results,
and their relation to theory.

\section{History}
\label{sec:history}

This field grew from studies of inclusive charmonium production
$\epem \to \psi^{(\prime)}\, X$~\cite{psiX-belle,psiX-babar}. 
An old CLEO analysis on a small sample~\cite{psiX-cleo} had presented evidence
for direct decays $\ufours\to J/\psi\,X$,
distinguished by $J/\psi$ momenta above the endpoint for 
$\ufours\to\BBbar[\to J/\psi\,X]$. 
Using data from early B-factory
running---$29.4\,\ifb$ on the \ufours\ resonance,
and $3.0\,\ifb$ in the continuum $60\,\mev$ below---Belle~\cite{psiX-belle}
excluded such production, setting a limit
$\br(\ufours\to J/\psi\,X) < 1.9\times 10^{-4}$ at 95\% confidence.
More importantly, they established picobarn cross-sections for
$\epem\to\psi^{(\prime)}\,X$ processes in the continuum,
and a peculiar momentum spectrum for the produced $\psi^{(\prime)}$:
in the $J/\psi$ sample, the cross-section fell to zero well below the
momentum endpoint (see Fig.~3 of Ref.~\cite{psiX-belle}).

\begin{figure}
  \includegraphics[width=8cm]{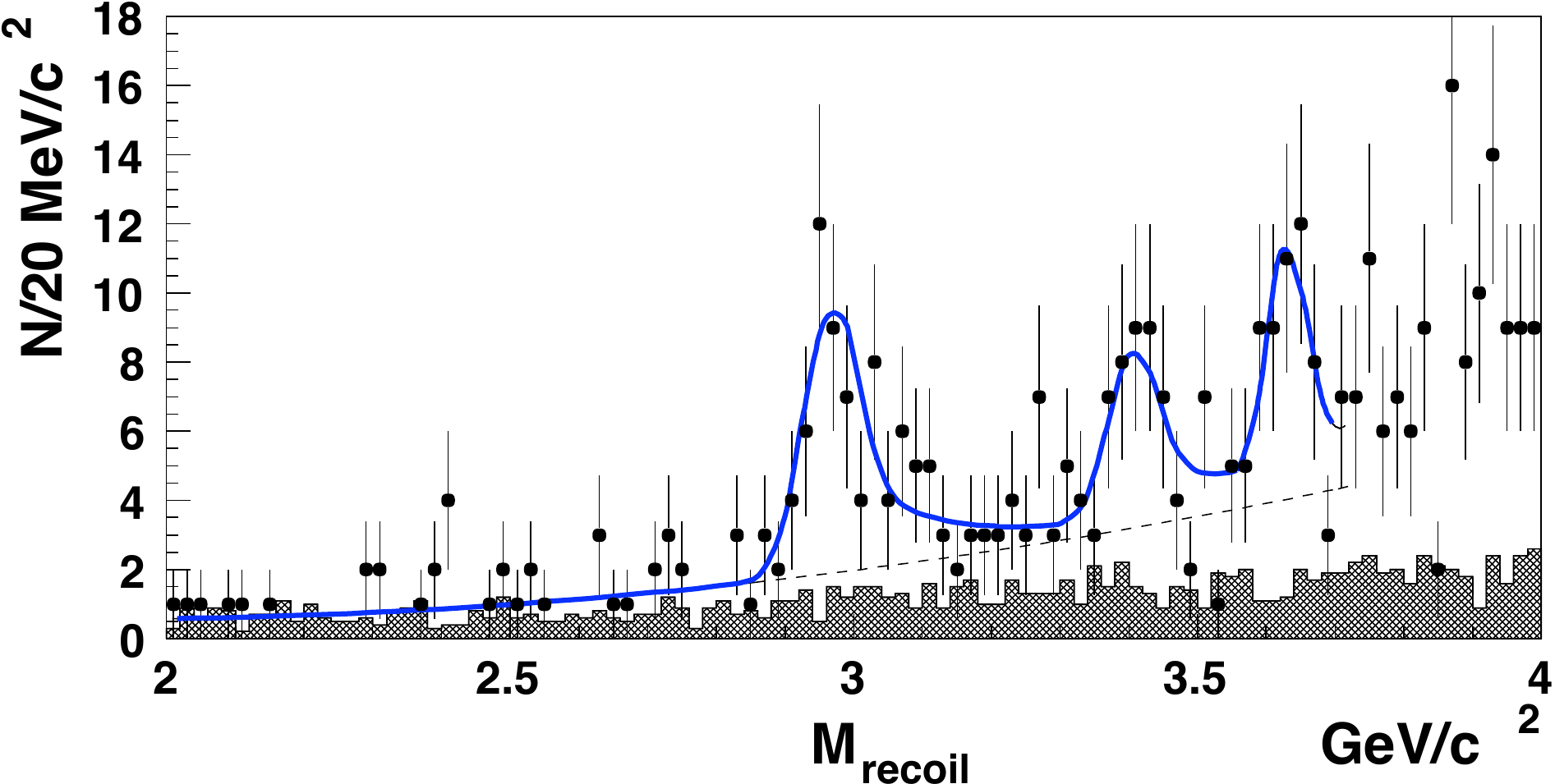}
  \caption{The mass spectrum of the
	unreconstructed (recoiling) system $X$ in $\epem\to J/\psi\,X$
	production in early Belle data~\cite{psiccbar-belle}.
	The curves show a fit to a series of charmonium
	states (solid) over a continuous component (dashed);
	the shaded histogram shows the distribution for events in the
	$J/\psi$ mass sidebands. The charmonium lineshapes are broadened by
	initial state radiation: see the text.}
  \label{fig:psiccbar-belle}
\end{figure}

The \epem\ initial state being known,
a simple rescaling of the $J/\psi$ momentum gives
the mass of the system recoiling against it in the final state:
\begin{equation}
  \mrecoil(\psi)=\sqrt{(\sqrt{s} - E^*_\psi)^2 - (p^*_\psi)^2}
  \label{eq:mrecoil}
\end{equation}
An upper bound on momentum thus corresponds to a \emph{lower bound} on the
mass of the recoiling system.
Studied in this way~\cite{psiccbar-belle},
the surprising conclusion (Fig.~\ref{fig:psiccbar-belle}) was that
the interaction $\epem\to J/\psi\,X$ does not proceed if the mass
of the recoiling system $X$ is below \ccbar\ threshold, while immediately above
threshold there is significant two-body production,
with $X = \eta_c$, $\chi_{c0}$, $\eta_c^\prime$.
Although familiar now, it is worth remembering how completely unexpected 
this result was at the time.

It was then accepted~\cite{nrqcd-old} that charmonium production
at $\sqrt{s} \approx 10.6\,\gev$ was dominated by
the $\epem \to \psi\, gg$ process, 
with a momentum spectrum extending to the kinematic endpoint
and thus a \mrecoil\ spectrum to low masses;
$\epem \to \psi\, \ccbar$ was an $\mathcal{O}(10\%)$ correction.
An additional contribution from $\epem \to \psi\, g$,
where the charmonium develops from a colour-octet \ccbar\ pair,
was expected at momentum endpoint (low \mrecoil).
The contradiction with data
led to reexamination of both theoretical and experimental methods.

In the recoil mass technique, the system $X$ is not reconstructed
and identification is thus indirect.
Alternative explanations of the data have sprung from this
limitation, while improvements to the method have focussed on 
partial reconstruction and constraints. 
The original analysis~\cite{psiccbar-belle} already used a mass-vertex
constraint in $J/\psi$ reconstruction, improving \mrecoil\ resolution
by a factor of two. Contributions from QED processes are more troublesome. 
Initial-state radiation (ISR) leads to a high-\mrecoil\ tail
(typically model-dependent) on the lineshape of any peak;
more than four tracks are required to suppress contributions from
other (low-multiplicity) QED interactions.

The leading alternative interpretation of the data relied on such a process:
$\epem\to \gamma^*\gamma^* \to \psi\,X$~\cite{twogamma-blb}.
While $\epem\to\gamma^*\to\psi\,X$ production requires the second state
to be even under charge conjugation, $\xi_C^X = +1$,
the two-virtual-photon process allows $\xi_C^X = \pm 1$ and in particular
permits $\epem\to\gamma^*\gamma^*\to J/\psi\,J/\psi$.
As only the $J/\psi\,\eta_c$ signal was statistically significant in the
2002 analysis~\cite{psiccbar-belle}, it was attractive to ascribe the peak
(in part) to events with a second $J/\psi$.
A more ambitious (and essentially mirror-image) proposal~\cite{glueball-bgl}
was that $\epem\to\psi\,gg$ did in fact dominate charmonium production,
with the gluons sometimes coupling to a glueball state
close to the $\eta_c$ mass.

These and other interpretations were effectively ruled out by 
an updated Belle analysis~\cite{psiccbar-belle-prd}, using $155\,\ifb$ of data
and techniques designed to discriminate between the theoretical options.
Three $\epem\to \psi\,\eta_c$ events were fully reconstructed, to be compared
with the $2.6\pm 0.8$ expected from the inclusive yield.
The recoil mass scale was also calibrated
using $\epem\to\gamma_{ISR}\,\psi'$ events and found to have a bias of less than
$3\,\mev$, ruling out confusion between $\eta_c$ and $J/\psi$.
Attempts to include $\epem\to\psi\,\psi^{(\prime)}$ components in the recoil
mass fit resulted in negative yields, and restrictive upper limits on 
their contribution (see Fig.~\ref{fig:psiccbar-both}, upper plot).
An eventual confirmation by BaBar~\cite{psiccbar-babar}
found comparable results (lower plot).

\begin{table}[b]
  \renewcommand{\arraystretch}{1.8}
  \caption{Coefficients from fits
	of the function $1+\alpha\cos^2\theta$ to $J/\psi$
	production (\tprod) and helicity angle (\thel) data
	at Belle~\cite{psiccbar-belle},
	for $\epem\to J/\psi\,\ccres$.
	See the text for results under the constraint 
	$\ahel=\aprod$, and the expectation.}
  \label{tab:angular}
  \begin{tabular}{lcccc}				\hline\hline
  \ccres	& \aprod\ 				
		& \ahel\  
		& $\ahel \equiv \aprod$
		&  expectation				\\ \hline
  $\eta_c$	& $+1.4^{+1.1}_{-0.8}$
		& $+0.5^{+0.7}_{-0.5}$
		& $+0.93^{+0.57}_{-0.47}$
		& $+1$ (P)				\\
  $\chi_{c0}$	& $-1.7\pm0.5$
		& $-0.7^{+0.7}_{-0.5}$
		& $-1.01^{+0.38}_{-0.33}$
		& $-1$ (S)				\\
$\eta_c^\prime$	& $+1.9^{+2.0}_{-1.2}$ 
		& $+0.3^{+1.0}_{-0.7}$
		& $+0.87^{+0.86}_{-0.63}$
		& $+1$ (P)				\\
					\hline\hline
  \end{tabular}
\end{table}

Belle also performed an angular analysis of $J/\psi\,\eta_c^{(\prime)}$
and $J/\psi\,\chi_{c0}$ events.
The distinctive forward peak in $J/\psi$ production angle ($\cos\tprod\to 1$)
predicted by the two-virtual-photon model 
for $J/\psi\,J/\psi$ (Fig.~2 of Ref.~\cite{twogamma-blb})
was not seen in data: results for all three states were consistent with 
shapes $1+\alpha\cos^2\theta$, and 
equal coefficients for production- and helicity-angle distributions
(see Table~\ref{tab:angular}) 
as expected for a single virtual photon.
The $J/\psi\,\eta_c^{(\prime)}$ fits were 
consistent with $\alpha=+1$ (P-wave production)
as required by $\eta_c$ quantum numbers,
and strongly disfavoured the $-0.87$ expectation for a
spin-zero glueball~\cite{glueball-bgl}. 
One- rather than two-virtual photon production (or the glueball explanation)
is thus favoured by all experimental tests.

\begin{figure}
  \begin{flushright}
    \includegraphics[width=8.0cm]{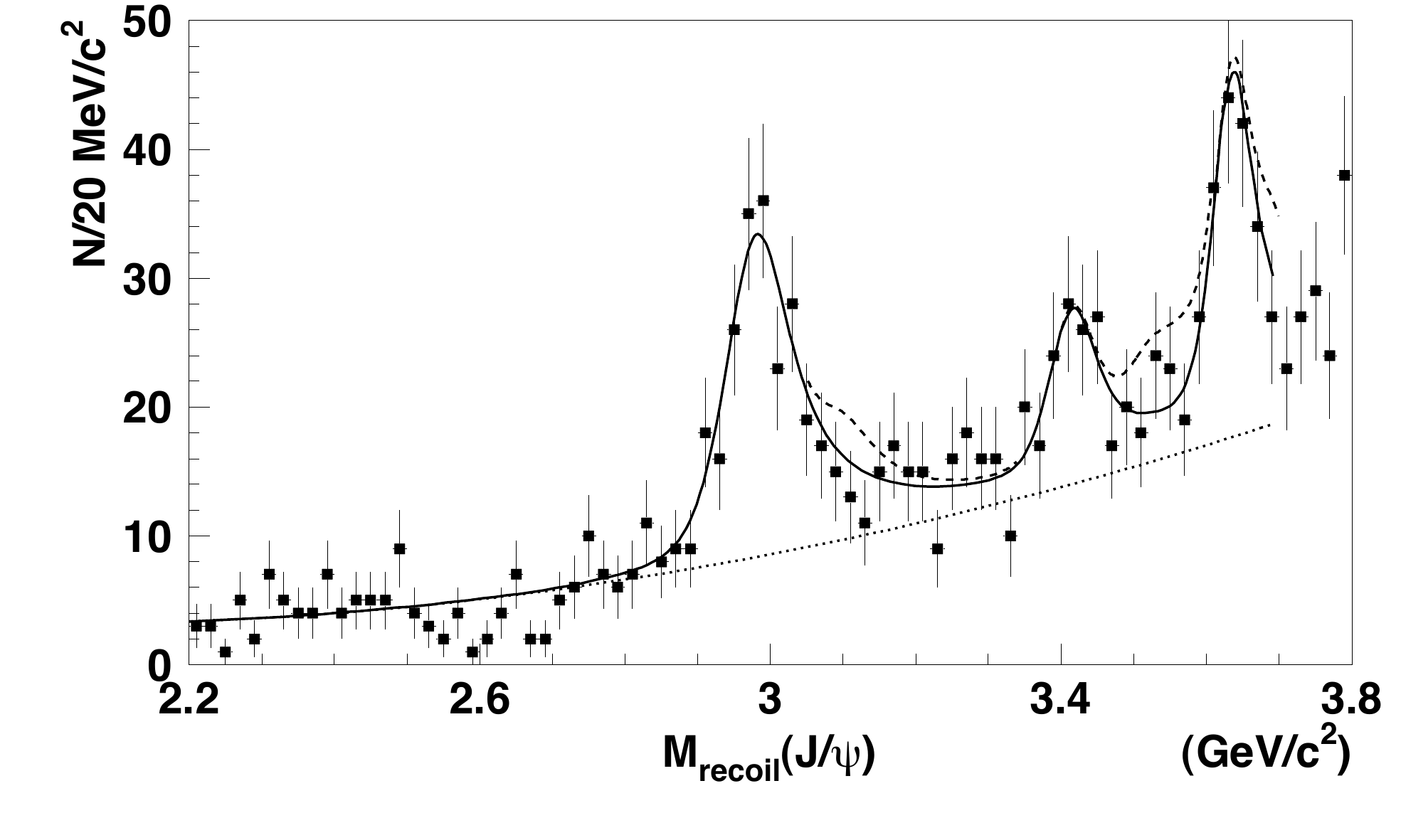} \\[0.5ex]
    \includegraphics[width=9.0cm]{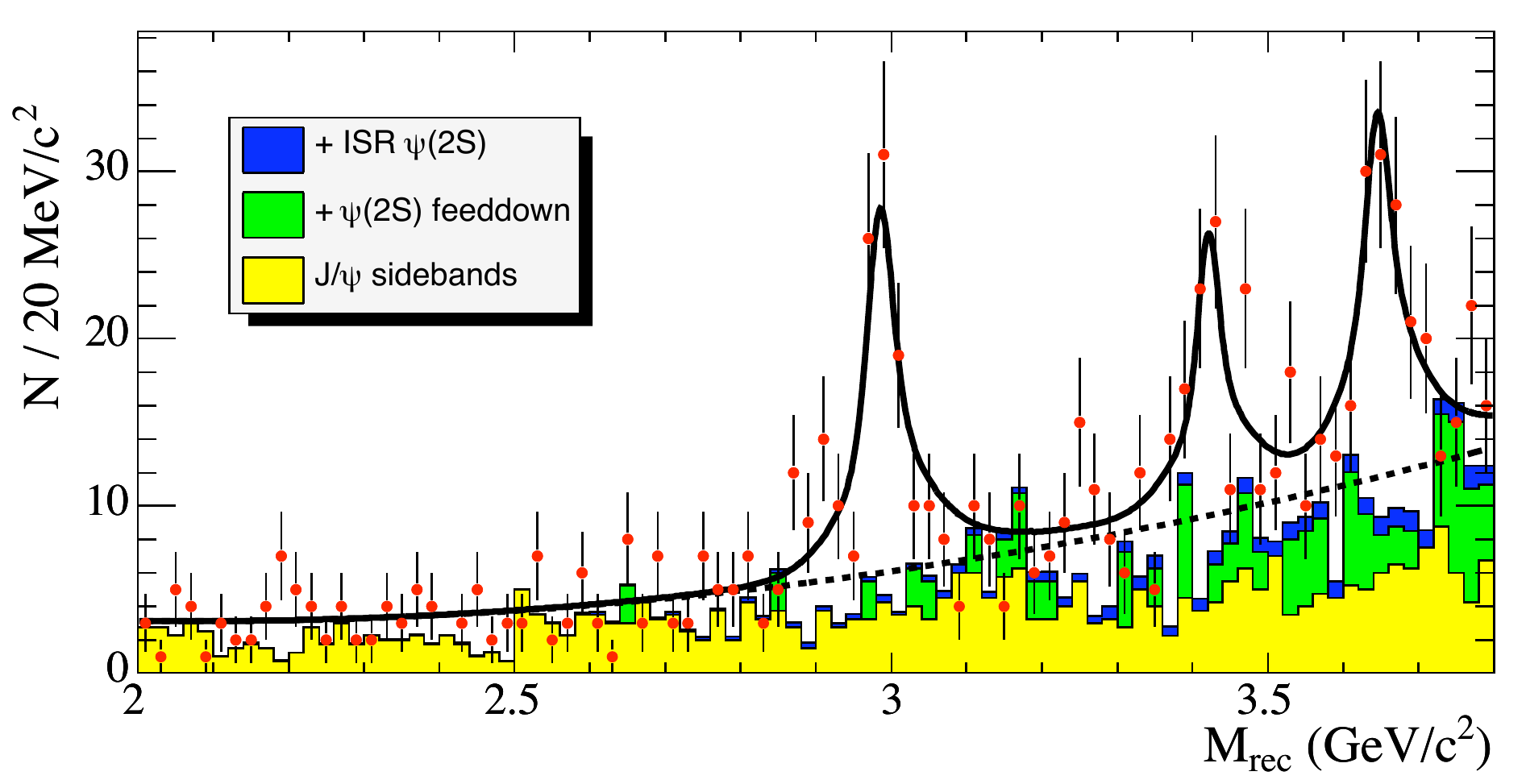}
  \end{flushright}
  \caption{Baseline $\epem\to J/\psi\,\ccres$ results from
	Belle (\cite{psiccbar-belle-prd}, upper plot),
	and the confirmation from BaBar (\cite{psiccbar-babar}, lower plot).
	In the upper plot the dashed line shows the upper limit contribution
	from final states where the fitted yield is insignificant or negative:
	$J/\psi\,(J/\psi,\,\chi_{c1,c2},\,\psi'$).}
  \label{fig:psiccbar-both}
\end{figure}


\section{Baseline results}
\label{sec:baseline}

\begin{table*}
  \caption{Double-charmonium production cross-sections from Belle and BaBar,
	with theoretical pre- and post-dictions.
	Due to background-suppression criteria, the experiments report effective cross-sections
	for the case where the unreconstructed state \ccres\ decays to at least 2 (``$>0$'')
	or 4 (``$>2$'') charged tracks, and thus underestimate
	the cross-section.}
  \label{tab:psiccbar-xsec}
      \renewcommand{\arraystretch}{1.4}
      \begin{tabular}{lr@{\hspace*{1cm}}c@{\hspace*{1cm}}c@{\hspace*{1cm}}c}											\hline \hline
				&				& \multicolumn{3}{c}{\ccres}					\\ \cline{3-5}
	\multicolumn{2}{l}{$\sigma(\epem\to\psi(nS)\,\ccres)$ [fb]}
				& $\eta_c(1S)$			& $\chi_{c0}$			& $\eta_c(2S)$			\\ \hline
	$\psi(2S)$, $\times \br_{>0}$
		& Belle~\cite{psiccbar-belle-prd}
				& $16.3 \pm4.6 \pm3.9$		& $12.5 \pm3.8 \pm3.1$		&  $16.0 \pm5.1 \pm3.8$		\\ \hline
	$\psi(1S)$, $\times\br_{>2}$  
		& Belle~\cite{psiccbar-belle-prd}
				& $25.6 \pm2.8 \pm3.4$		& $ 6.4 \pm1.7 \pm1.0$		&  $16.5 \pm3.0 \pm2.4$		\\
		&\textsc{BaBar}~\cite{psiccbar-babar}
				& $17.6\pm2.8^{+1.5}_{-2.1}$	& $10.3\pm2.5^{+1.4}_{-1.8}$	& $16.4\pm3.7^{+2.4}_{-3.0}$	\\ \hline
	$\psi(1S)$  
		& Braaten and Lee~\cite{nrqcd-braaten-lee}
				& $3.78\pm1.26$			& $2.40\pm1.02$			&  $1.57\pm0.52$		\\
		& \ldots with relativistic corrections~\cite{nrqcd-braaten-lee}
				& $7.4^{+10.9}_{-4.1}$		& --				& $7.6^{+11.8}_{-4.1}$		\\
		& Liu, He, and Chao~\cite{nrqcd-liu-he-chao}
				& $5.5$                 	& $6.9$				&  $3.7$			\\
		& Zhang, Gao, and Chao~\cite{nrqcd-zhang-gao-chao}
				& $14.1$			& --				&  --				\\ \hline
		& Bondar and Chernyak~\cite{lc-bondar-chernyak}
				& $33$				& --				& --				\\ \hline\hline
      \end{tabular}
\end{table*}

Belle~\cite{psiccbar-belle-prd} and BaBar~\cite{psiccbar-babar} measurements
of $\epem \to \psi^{(\prime)}\, \ccres$ cross-sections are summarised in
Table~\ref{tab:psiccbar-xsec}: with competing explanations excluded,
and reasonable agreement between the two experiments, 
these results are no longer in serious dispute.
Remarkably, there seems to be no suppression of radially-excited states:
cross-sections for
$\psi\,\eta_c$, $\psi\,\eta_c^\prime$,
$\psi'\,\eta_c$, and $\psi'\,\eta_c^\prime$
are all comparable.
This presumably contains some hint as to the production mechanism.

Low-order perturbative calculations,
as embodied in nonrelativistic quantum mechanics
(NRQCD,~\cite{nrqcd-braaten-lee,nrqcd-liu-he-chao,nrqcd-zhang-gao-chao})
underestimate the cross-sections by an order of magnitude or more;
the discrepancy is reduced, but not removed,
when relativistic corrections are taken into account~\cite{nrqcd-braaten-lee}.
Neither are other features of the data well-explained: 
NRQCD predicts $\alpha \simeq +0.25$
for $\epem\to J/\psi\,\chi_{c0}$~\cite{nrqcd-braaten-lee},
disfavoured by the Belle analysis (Table~\ref{tab:angular}),
which prefers pure S-wave production.

A calculation in the light-cone formalism~\cite{lc-bondar-chernyak},
however, appears to match at least the $J/\psi\,\eta_c$ cross-section.
A variety of theoretical approaches are currently being pursued,
and it is no longer easy to characterise the issues at stake:
on comparison of NRQCD and light-cone estimates, see for example the
very extended discussion in Ref.~\cite{nrqcd-lc-comparison}.
Analysis of the merits of this theoretical work is outside
the scope (and competence) of this review.\footnote{We note that
	a theoretical analysis has appeared
	since the workshop~\cite{nrqcd-latest}, claiming that
	the discrepancy in the $J/\psi\,\eta_c$ cross-section
	between NRQCD and experiment is now resolved.}
It does however seem clear that experimental work is still driving, 
rather than being directed by, theoretical studies.


\section[Cutting edge]{The new cutting edge: \\ states above open-charm threshold}
\label{sec:cutting}

Active experimental work has now shifted to the case where the system recoiling
against the $\psi$ is above open-charm threshold.
Even allowing for the various
$\epem\to\dmes^{(*)}\dbar{}^{(*)}$ continuum components in fitting the 
inclusive $\mrecoil(\psi)$ spectrum, Belle found a 
$5.0\sigma$ peak at $(3936\pm14)\,\mev$ (Fig.~1 of Ref.~\cite{x3940-belle}).
The background under this peak (called $X(3940)$) being too large
for further detailed study, Belle explicitly reconstructed a \dmes-meson and 
then considered cases where the remaining system was close
to the \dmes\ or \dstar\ in mass: 
$\mrecoil(\psi\,\dmes) \approx m_{\dmes^{(*)}}$.
Constraining such cases to match the $\dmes^{(*)}$ mass
also improved the $\mrecoil(\psi)$ resolution.
This allowed the reconstruction of a clear $X(3940)\to\dstar\dbar$ peak,
and an upper limit on the same structure
in $\dmes\dbar$ (Fig.~3 of~\cite{x3940-belle})

\begin{figure}
  \includegraphics[width=7.5cm]{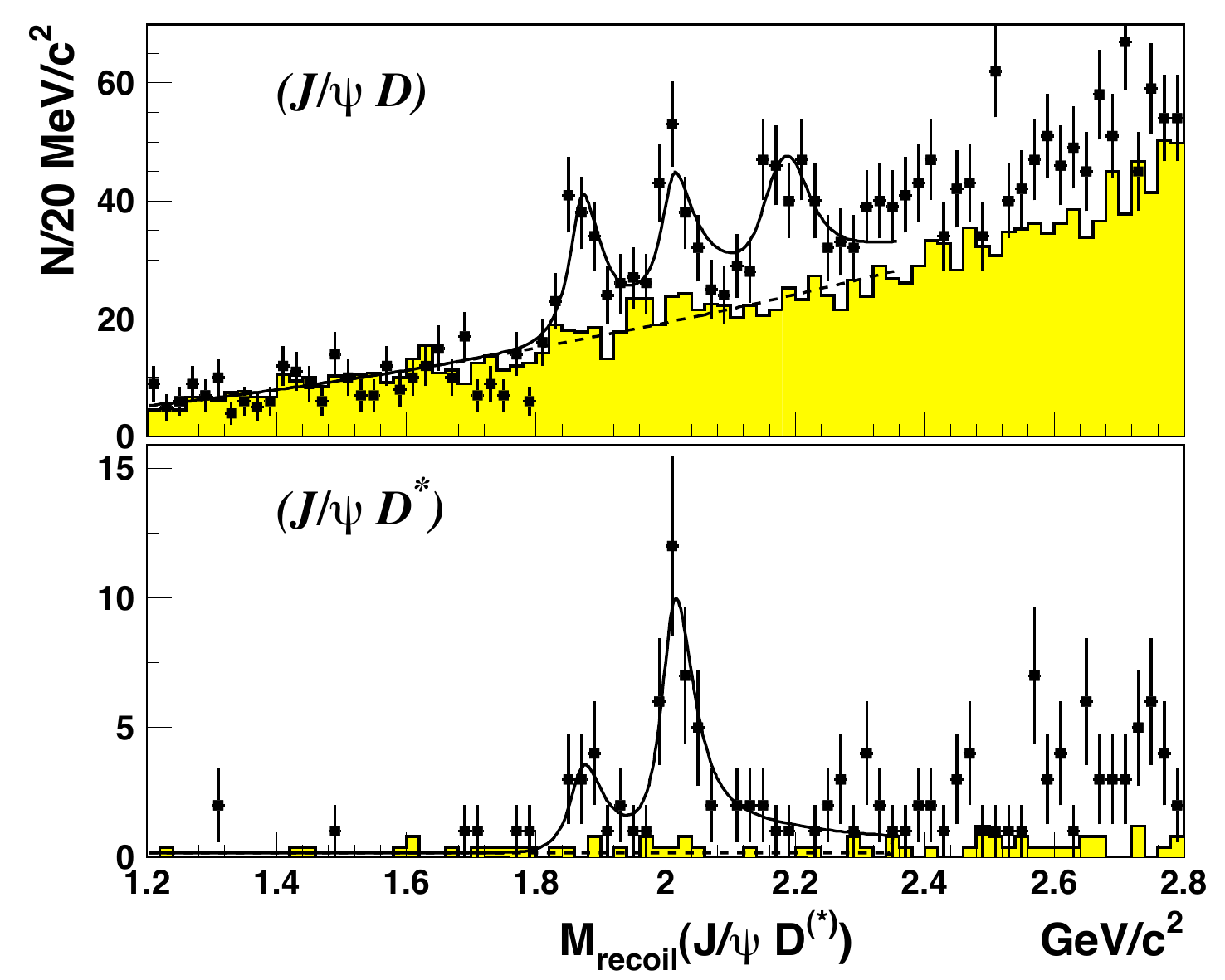}
  \caption{The mass of the system recoiling against fully-reconstructed
	$J/\psi$ and \dmes\ (upper plot) or \dstarp\ (lower plot) mesons
	at Belle~\cite{x3940-belle-update}.
	The shaded histogram shows the distribution in $\dmes^{(*)}$-mass
	sidebands: the fitted curve is discussed in the text.}
  \label{fig:x3940-dtag-data}
\end{figure}

Belle has released an updated analysis~\cite{x3940-belle-update}
based on systematic use of this $\dmes^{(\ast)}$-tagging technique.
Fig.~\ref{fig:x3940-dtag-data} shows the $\mrecoil(\psi\,\dmes^{(*)})$ spectrum
for $693\,\ifb$ of data, after reconstruction and mass-constraint 
of the $J/\psi$, and then a \dz, \dplus, or \dstarp\ meson.
A simultaneous fit with the $\dmes^{(*)}$-mass sidebands is performed: clear
and significant peaks are seen, corresponding to processes
$\epem\to\psi\,\dmes\dbar$, $\psi\,\dstar\dbar$, and $\psi\,\dstar\dbar{}^*$.

Monte Carlo study shows that we indeed expect these processes to be
reconstructed in this way, with recoil mass resolution of about $30\,\mev$:
smaller than the difference in \dmes\ and \dstar\ masses.
Disjoint samples
$\left|\mrecoil(\psi\,\dmes{}^{(*)})-m_{\text{tag}}\right| < 70\,\mev$ where
the unreconstructed system is tagged as a \dmes\ or a \dstar\ are thus selected
(ISR leads to a 10\% $\psi\dmes\dbar \to \psi\dmes\dbar{}^*$ cross-feed),
and $\mrecoil(\psi\,\dmes{}^{(*)})$ is then constrained to the mass of the 
tagged meson. This improves the resolution on $M(\dmes^{(*)}\dbar{}^{(*)})$
by a factor of 3--10: results for the three samples are shown in 
Fig.~\ref{fig:x3940-dd}.
Peaks above the background are seen near threshold in each sample.

\begin{figure}
  \includegraphics[width=6.5cm]{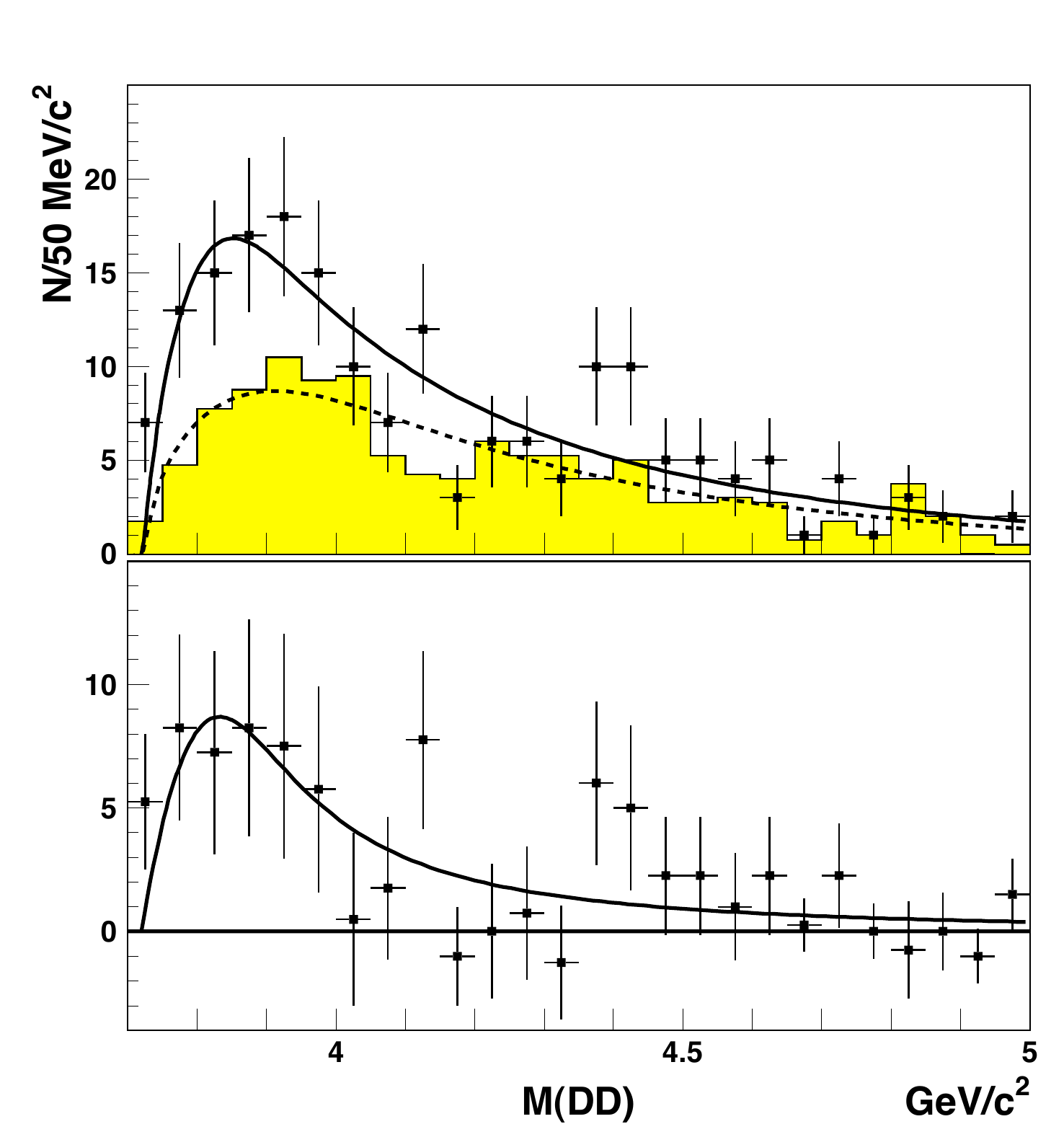}
  \\[0.5cm]
  \includegraphics[width=6.5cm]{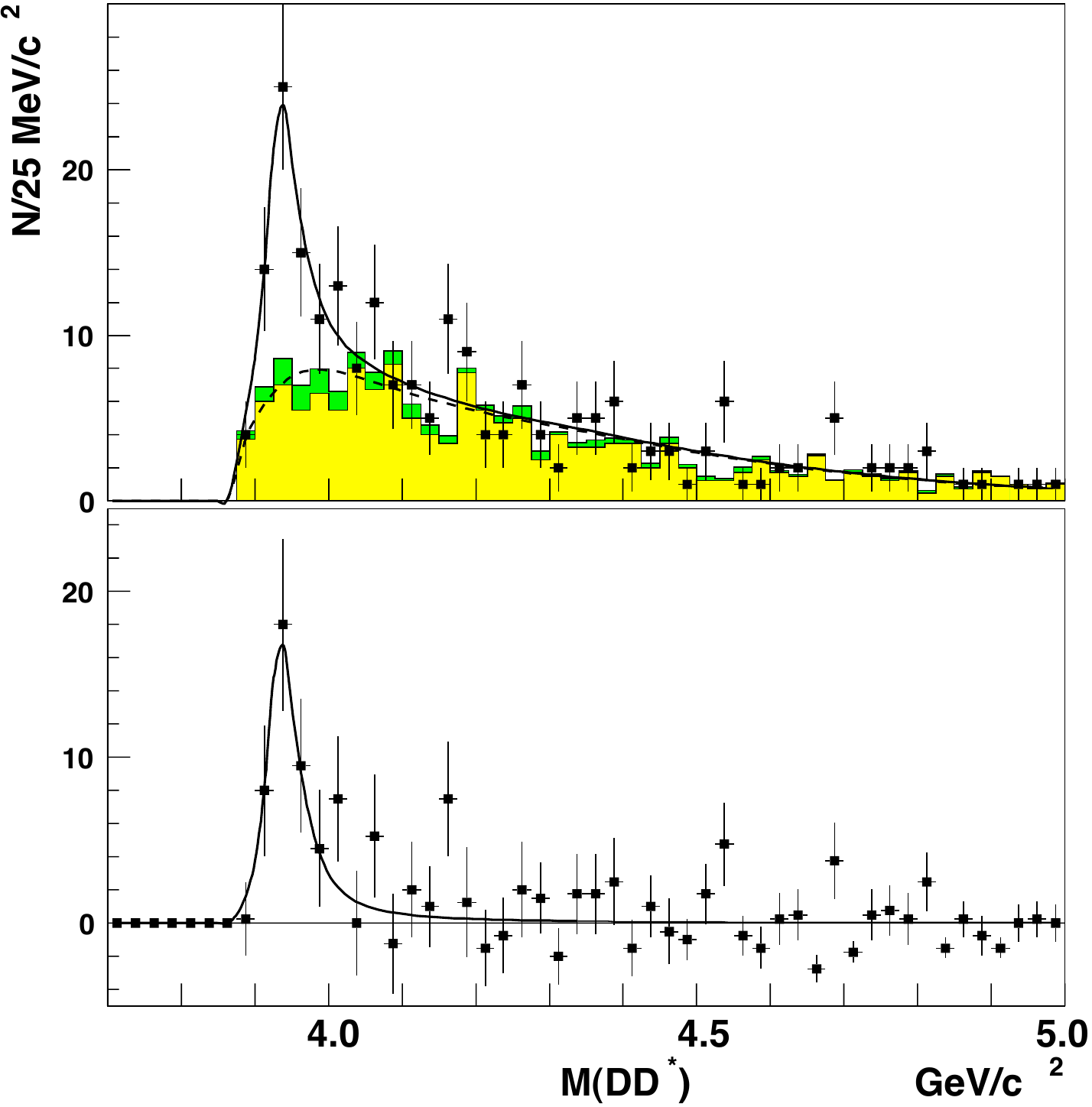}
  \\[0.5cm]
  \includegraphics[width=6.5cm]{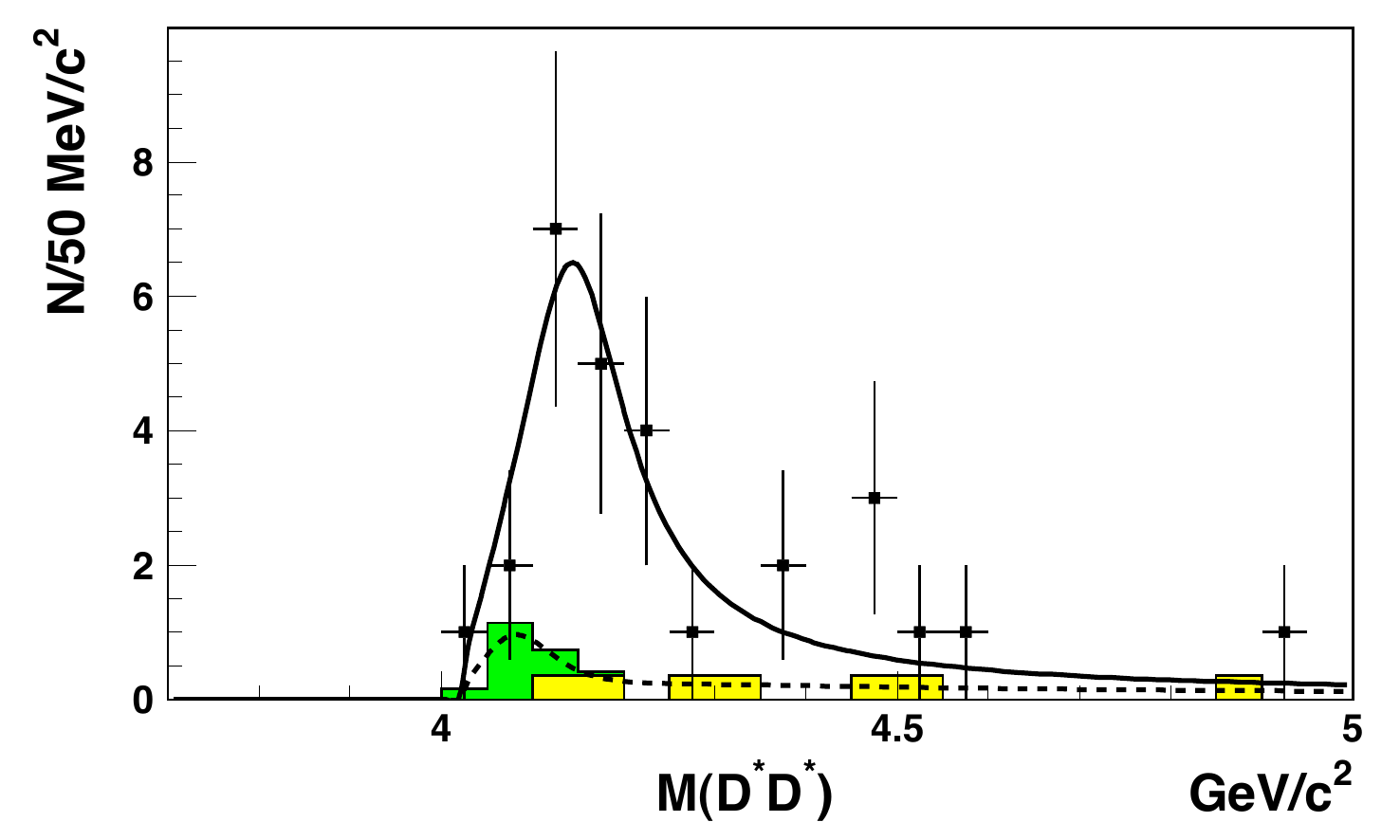}
  \caption{$M(\dmes^{(*)}\dbar{}^{(*)})$ spectra
	for $\epem\to J/\psi\,\dmes^{(*)}\dbar{}^{(*)}$ events
	at Belle~\cite{x3940-belle-update},
	where the first \dmes-meson is explicitly reconstructed,
	and the associated \dmes-meson is recovered via recoil mass selections
	and constraints: (upper) $\dmes\dbar$, (middle) $\dmes\dstarb$,
	and (lower) $\dstar\dstarb$ samples are shown.
	Distributions in the reconstructed $\dmes^{(*)}$-mass sidebands
	are shown in yellow,
	and reflections (middle) $\dmes\dbar \to \dmes\dstarb$ 
	and (lower) $\dmes\dstarb \to \dstar\dstarb$ in green.
	The solid curves show the fits described in the text, 
	and the dashed curves their combinatorial background and reflection
	components. 
	The second panel in the upper and middle plots shows the data and fit
	with these components subtracted.}
  \label{fig:x3940-dd}
\end{figure}

Combinatorial backgrounds are taken into account via simultaneous fits to
the data in the reconstructed $\dmes^{(*)}$-mass signal and sideband regions.
The excess over background is fitted with the sum of a threshold function
to represent non-resonant $\epem\to\psi\,\dmes^{(*)}\dbar{}^{(*)}$ production,
and an S-wave relativistic Breit-Wigner. In all cases the threshold term is
insignificant, and a significant Breit-Wigner peak is seen.
In $M(\dmes\dbar)$ the peak is broad, and the fit unstable against variations
of its conditions; in $M(\dmes\dstarb)$ and $M(\dstar\dstarb)$ the fit is stable
and the resonant peak significant at over $5\sigma$.
Various cross-checks are performed, including Monte Carlo and data tests of
the background shape in $\dmes^{(*)}$-mass sideband and signal regions; 
study of charged- and neutral-\dmes\ subsamples;
and fitting of events with reconstructed \dstar, and associated \dbar:
the latter obtains results consistent with the $J/\psi\,\dmes\dstarb$ analysis of 
Fig.~\ref{fig:x3940-dd} (middle), but with lower efficiency and significance.
Parameters for the resonant enhancements are summarised
in Table~\ref{tab:x-ddbar-param}.

\begin{table}
  \caption{Resonant enhancements in $\epem\to\psi\,\dmes^{(*)}\dbar{}^{(*)}$ production
	at Belle~\cite{x3940-belle-update}. The $\psi$ and the meson shown are fully
	reconstructed and then refitted under a mass constraint: the remaining meson is
	identified as discussed in the text.}
  \label{tab:x-ddbar-param}
  \renewcommand{\arraystretch}{1.4}
  \begin{tabular}{lccc}									\hline\hline
	Final state	& \dmes\dbar		& \dmes\dstarb		& \dstar\dstarb		\\
	Reconstructed	& \dmes			& \dmes			& \dstar		\\
	Resonant term	& ---			& $X(3940)$		& $X(4160)$		\\\hline
	Mass (\mev)	& $3878\pm48$		& $3942^{+7}_{-6}\pm6$	&$4156^{+25}_{-20}\pm15$\\
	Width (\mev)	& $347^{+316}_{-143}$	& $37^{+26}_{-15}\pm8$	&$139^{+111}_{-61}\pm21$\\
	Significance	& $4.4\sigma$ 		& $6.0\sigma$ 		& $5.5\sigma$		\\
	Fit behaviour	& unstable		& stable		& stable		\\\hline
	$\sigma(\epem\to\psi\,X)$		&		&				\\
	$\times \br_{\dmes^{(*)}\dbar{}^{(*)}}$ (fb)
			& ---			& $13.9^{+6.4}_{-4.1}\pm2.2$
									& $24.7^{+12.8}_{-8.3}\pm5.0$
												\\
											\hline\hline
  \end{tabular}
\end{table}

The $X(4160)$ enhancement has not previously been reported.
The $X(3940)$ mass and yield results are consistent with those of
the earlier analysis~\cite{x3940-belle},
while the width is larger than the published value
of $(15.1\pm10.1)\,\mev$: a likelihood function non-parabolic
in the width parameter had been noted in that case,
with a $52\,\mev$ upper limit at 90\% confidence.
The corresponding limit in the new analysis is
$\Gamma< 76\, \mev$.
  
Cross-sections for $J/\psi\,X(3940)$ and $J/\psi\,X(4160)$ production 
(Table~\ref{tab:x-ddbar-param}, last row) are in the 20 femtobarn class,
as for all the significant $\epem\to\psi^{(\prime)}\,\ccres$ processes seen to date.


\section{Sidelines}
\label{sec:sidelines}

The concentration on experimentally fruitful problems --- establishing
and measuring quasi-two-body processes $\epem\to\psi^{(\prime)}\,\ccres$ ---
has led to a relative neglect of inclusive $\epem\to\psi^{(\prime)}X$ studies.
The 2002 Belle analysis~\cite{psiccbar-belle} established the fraction
\begin{equation}
  \frac{\sigma(\epem\to\psi\,\ccbar)}{\sigma(\epem\to\psi\,X)}
  = 0.59^{+0.15}_{-0.13}\pm0.12
  \label{eq-ccbar-fraction}
\end{equation}
using the following method:
reconstruction and mass-constraint of $J/\psi$, 
and then an associated $\dmes^{(*)}$-meson;
rejection of contamination from $\epem\to\BBbar$ events
using momentum requirements; 
and a two-dimensional fit to obtain $\psi\,\dmes^{(*)}X$ yields
(see Fig.~2 of Ref.~\cite{psiccbar-belle}).
The cross-section thus obtained is model-dependent,
relying on simulated $\ccbar \to \dmes^{(*)}X$ fragmentation
by PYTHIA to establish efficiencies.
There has been no published update of this remarkable result.

(Model-independent but since-unpublished Belle results were presented
to the Quarkonium Working Group in 2002,
relying instead on counting of \dz, \dplus, and all ground-state charmed
hadrons under minimal cuts. Resolution was similar, with a lower limit on the 
$J/\psi\,\ccbar$ fraction of 0.48 at 95\% confidence.)

\begin{figure}
  \includegraphics[width=7.5cm]{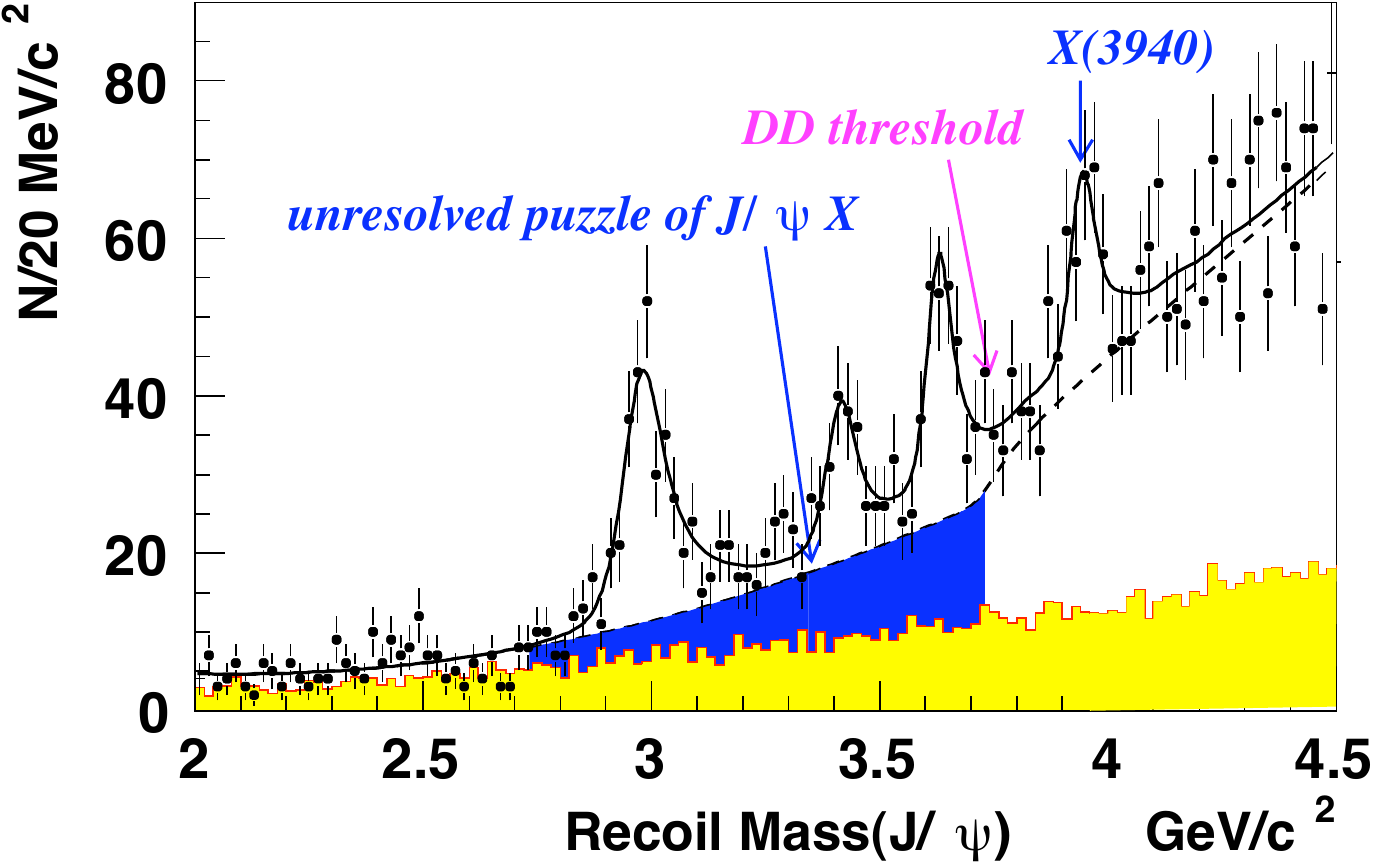}
  \caption{Inclusive $\epem\to J/\psi\,X$ data, showing the
	combinatorial background (yellow)
	and non-$J/\psi\,\ccres$ contributions (blue)
	below open-charm threshold.
	(Adapted from a figure by P.~Pakhlov.)}
  \label{fig:x3940-pakhlov-conf}
\end{figure}

A complementary problem is the nature of $\epem\to J/\psi\,X$ production 
by processes other than $J/\psi\,\ccbar$. 
Such production does seem to occur: 
there is a continuous component in inclusive recoil mass spectra
(Fig.~\ref{fig:x3940-pakhlov-conf}) in excess of 
background, below the open-charm threshold. 
The mystery is why this component --- due to the 
$\epem\to\psi\,g$ process? --- should obey the \ccbar\ threshold.
A problem for experimental study here is the lack of theoretical guidance:
predictions for the various $\epem\to\psi\,X$ processes that so preoccupied
the original Belle~\cite{psiX-belle} and BaBar~\cite{psiX-babar} analyses
have been discredited, but new predictive studies have not taken their place.
Such work is overdue.


\section{Summary}
\label{sec:summary}

The period of fundamental doubt about double charmonium production is now over.
Questions concerning the experimental method have been addressed, and the
potential for confusion by $\epem\to\gamma^\ast\gamma^\ast\to J/\psi\,X$
or other more exotic processes has been excluded~\cite{psiccbar-belle-prd}.
And the fear that some untraceable mistake had been made is effectively
dispelled by BaBar's confirmation~\cite{psiccbar-babar}
of the Belle results~\cite{psiccbar-belle}.

Those results establish that $\epem\to\psi\,\ccbar$ dominates charmonium
production in the continuum at $\sqrt{s} \approx 10.58\,\gev$,
while $\psi^{(\prime)}\,\ccres$ cross-sections are at the $20\,\fb$ level,
with no suppression of radially excited states.
Recent work shows that prominent resonant contributions continue
above open-charm threshold, with similar cross-sections;
this process is proving fruitful in the search for new hidden-charm states.

The ball now lies in the court of theory.
Interpretive work on the new states $X(3940)$ and $X(4160)$ is already underway,
but a predictive account of $\epem\to\psi^{(\prime)}\,X$ amplitudes is still lacking.
An advantage of the NRQCD approach is its pretension to universal application:
at the Tevatron, for example, and the LHC.
We await an accurate account of double-charmonium production
in \epem\ annihilation that can also embrace quarkonium production
at other facilities.


\end{document}